\documentclass[aps,pra,twocolumn,groupedaddress,showpacs]{revtex4}
\usepackage{amsmath,amsfonts,amssymb,bbm,citesort,graphicx}

\newcommand{\eqn}[1]{\begin{equation} #1 \end{equation}} 
\newcommand{\aln}[1]{\begin{align} #1 \end{align}}       
\newcommand{\mc}{\mathcal}                               
\newcommand{\mbf}{\mathbf}                               
\newcommand{\bs}{\boldsymbol}                
\newcommand{\eq}[1]{(\ref{#1})}              
\newcommand{\wbar}{\overline}                
\renewcommand{\l}{\left}                     
\renewcommand{\r}{\right}                    

\begin{document}


\title{Mode statistics in random lasers}


\author{Oleg Zaitsev}
\email[E-mail: ]{oleg.zaitsev@uni-duisburg-essen.de}
\affiliation{Fachbereich Physik, Universit\"at Duisburg-Essen, 
             Lotharstr.~1, 47048 Duisburg, Germany}



\begin{abstract}

Representing an ensemble of random lasers with an ensemble of random matrices,
we compute average number of lasing modes and its fluctuations. The regimes of
weak and strong coupling of the passive resonator to environment are
considered. In the latter case, contrary to an earlier claim in the
literature, we do not find a power-law dependence of the average mode number
on the pump strength. For the relative fluctuations, however, a power law can
be established. It is shown that, due to the mode competition, the
distribution of the number of excited modes over an ensemble of lasers is not
binomial. 

\end{abstract}


\maketitle

\section{Introduction}

Random lasers with coherent feedback (see Ref.~\cite{cao03} for a review) are
systems based on active disordered materials where self-sustained radiation
modes can be formed. Another possibility involves sufficiently open
wave-chaotic resonators filled with active media. In the absence of pumping,
both realizations are characterized by short-lived strongly interacting
passive modes. In order to treat such systems, the standard laser
theory~\cite{sarg74,hake85}, which assumes an almost closed resonator, needs
to be modified. The first step in this direction was made in
Ref.~\cite{hack02}, where the Langevin formalism was adapted to the present
situation. The authors described lasing-mode oscillations by non-Hermitian
matrices and derived an expression for the linewidth. In Ref.~\cite{hack03} a
connection was made between the Langevin and the master equations. An improved
treatment of non-linearities in the multimode laser theory was proposed in
Ref.~\cite{ture06}. 

An interesting problem (also from experimentalist's standpoint) is the effect
of mode competition on the number of lasing modes. Equations yielding this
number were derived in Refs.~\cite{hake63,misi98} for a weakly open resonator.
The authors of Ref.~\cite{misi98} discovered that the average number of modes
varies (within certain limits) as a power of the pump strength. The theory is
based on the fact that the passive-mode widths for a weakly open resonator
follow a $\chi^2$~distribution. The results were tested numerically by
generating ensembles of possible widths according to this distribution. In
Ref.~\cite{hack05} the mode-number equation of Ref.~\cite{misi98} was
rederived for a resonator with overlapping passive modes. In particular,
Ref.~\cite{hack05} exploits a possibility to reproduce statistical properties
of modes in open chaotic resonators with the specially chosen ensembles of
non-Hermitian random matrices. The average mode number was computed from sets
of passive-mode widths obtained directly from these ensembles. It was
claimed~\cite{hack05} that this number has a power-law dependence on pumping
in a resonator strongly coupled to the environment. The theory of
Ref.~\cite{hack05} relies on the eigenvector and eigenvalue statistics of
non-Hermitian random matrices, which was extensively studied in the
literature~\cite{fyod97,fyod99,somm99,fyod03}. However, general analytical
expressions for the relevant correlations of the left and right eigenvectors
are still unknown. The correlations were carefully studied numerically in
Ref.~\cite{hack05}. 

In the present paper we employ the approach of Ref.~\cite{hack05} to study the
average number of excited modes and, for the first time, its fluctuations. In
the weak-coupling case, our results for the average agree with the predictions
of Ref.~\cite{misi98}. For strong coupling, contrary to the results of
Ref.~\cite{hack05}, we do not find a power law for the average, but do find it
for the \emph{relative} fluctuations. Relating the average number of lasing
modes to its fluctuations, it is possible to extract some information about
the distribution of this number over an ensemble of lasers. We argue that a
binomial distribution gets distorted by the passive-mode overlap and the
active-mode competition.

\section{Theoretical background}

We begin by recalling the derivation of Eq.~\eq{N}~\cite{misi98,hack05}, which
yields the number of lasing modes for a given pump strength. This equation is
analyzed numerically in the subsequent sections.

\subsection{Langevin equations}

The system under consideration comprises an open resonator filled with $\mc N$
identical atoms. A random laser can be modeled by a resonator with an
irregular shape, such that its eigenfunctions are chaotic. The system
Hamiltonian
\eqn{
  H_{\text{sys}} = H_{\text{f}} + H_{\text{a}} + H_{\text{f-a}}
}
is a sum of the radiation-field Hamiltonian~$H_{\text f}$, atomic
Hamiltonian~$H_{\text a}$, and their interaction~$H_{\text{f-a}}$. 

It is convenient to represent the system as field and atoms in an ideal
(isolated) resonator interacting with environment (heat and pump reservoirs,
or baths). The reservoir degrees of freedom are then eliminated. The
reservoirs acting on the field and on the atoms are assumed to be independent
of each other~\cite{hake84}. Accordingly, the field Hamiltonian can be written
in the form
\eqn{
  H_{\text{f}} = \hbar \sum_\lambda \omega_\lambda\, a_\lambda^\dag a_\lambda + 
  \Delta H_{\text{f}},
}
where $a_\lambda$ are the annihilation operators for the modes of the closed
resonator with frequencies~$\omega_\lambda$ and $\Delta H_{\text{f}}$ includes
the bath Hamiltonian and the resonator-bath interaction. In the case of an
empty resonator with an opening, the role of the reservoir is played by an
external field having a continuous spectrum. This model, adopted also for the
present paper, was carefully studied in Refs.~\cite{vivi03,vivi04}, where, in
particular, the ways to split the field into the internal (resonator) and
external (bath) parts were discussed.

The atoms will be approximated by their two active levels separated by
energy~$\hbar \nu$. Given the Fermionic operators $c_{1,2}$ for the levels, one
can define the pseudospin-$1/2$ operators $\sigma \equiv s_x - i s_y = c_1^\dag
c_2$ and $s \equiv s_z = \frac 1 2 \l(c_2^\dag c_2 - c_1^\dag c_1 \r)$, i.e.,
$s_i$ generate a spin su(2) algebra. The spin operators $\sigma_p$ and $s_p$
($p = 1, \ldots, \mc N$) of different atoms commute. The atomic Hamiltonian
becomes 
\eqn{
  H_{\text{a}} = \hbar \nu \sum_{p=1}^{\mc N} s_p + \Delta H_{\text{a}},
}
where the reservoir operators are contained in~$\Delta H_{\text{a}}$.

The interaction Hamiltonian 
\eqn{
  H_{\text{f-a}} = i \hbar \sum_{\lambda p} \l( g_{\lambda p} a_\lambda^\dag
  \sigma_p - \text{h.c.} \r)
}
is written in the rotating-wave and dipole approximations. The former neglects
the terms proportional to~$a_\lambda \sigma_p$ and~$a_\lambda^\dag
\sigma_p^\dag$. Since $\nu \sim \omega_\lambda$, such antiresonant products
would oscillate with a double optical frequency in the interaction picture.
The latter approximation can be applied since the optical wavelength is much
larger than the atom size. Then the coupling constant (in the Gaussian units)
\eqn{
  g_{\lambda p} = \nu\, \sqrt{\frac {2 \pi} {\hbar \omega_\lambda}}\,
  \mbf d_{21} \cdot \bs \phi_\lambda^* (\mbf r_p)  
}
is expressed in terms of the dipole moment~$\mbf d_{21}$ for the $1 \to 2$
transition between the atomic levels, as well as the normalized vector-valued
eigenfunction~$\bs \phi_\lambda (\mbf r_p)$ of the mode~$\lambda$ at the atom
position~$\mbf r_p$~\cite{wall95}. In chaotic resonators the values of an
eigenfunction at any two positions are uncorrelated (apart from  normalization
and boundary effects) if they are more than a wavelength apart~\cite{berr77}.
Hence, the couplings~$g_{\lambda p}$ can be treated as independent Gaussian
random variables. 

In the Heisenberg picture, an equation of motion for an operator~$O$ is $\dot O
= \frac i \hbar \l[ H, O \r]$. For the laser operators $a_\lambda$, $\sigma_p$,
and $s_p$, the equations of motion can be cast in the form of Langevin
equations:
\aln{
  &\dot a_\lambda = -i \omega_\lambda a_\lambda - \sum_{\lambda'}
  \gamma_{\lambda \lambda'} a_{\lambda'} + \sum_{p'} g_{\lambda p'} \sigma_{p'}
  + F_\lambda, 
  \label{langlamb}\\
  &\dot \sigma_p = -\l( i \nu + \gamma_\perp \r) \sigma_p + 2 \sum_{\lambda'}
  g_{\lambda' p}^*\, a_{\lambda'} s_p + F^\sigma_p, 
  \label{langsigm}\\
  &\dot s_p = \gamma_\parallel \l(S - s_p \r) - \sum_{\lambda'} \l( g_{\lambda'
  p}\, a_{\lambda'}^\dag\, \sigma_p + \text{h.c.} \r) + F^s_p.
  \label{langs}
}
The reservoirs enter the equations via the damping ($\gamma_{\lambda
\lambda'}$, $\gamma_\perp$, $\gamma_\parallel$) and pumping ($S$) parameters
and the operators of stochastic forces ($F_\lambda$, $F^\sigma_p$, $F^s_p$).
The latter have zero reservoir average and are $\delta$-correlated in time.
This property is a consequence of the Markov approximation, which requires the
reservoir relaxation time to be much smaller than all the other time scales.
Equations \eq{langlamb}-\eq{langs} are appropriate for chaotic resonators. 
They differ from the standard equations of the laser theory~\cite{hake85} in
to aspects: the non-diagonality of~$\gamma_{\lambda \lambda'}$ and the
randomness of~$g_{\lambda p}$.  Equation~\eq{langlamb} was derived in
Refs.~\cite{vivi03,hack03}. The damping matrix~$\gamma_{\lambda \lambda'}$ is
Hermitian. It is strongly non-diagonal if the resonator modes are overlapping.
The off-diagonal elements point to the interaction between the respective
modes via a coupling to the continuum. Equations~\eq{langsigm} and~\eq{langs}
follow, e.g., from the Langevin theory for three-level atoms~\cite{sarg74} if
the total population of the two active levels is kept constant.
$\gamma_\perp$~and $\gamma_\parallel$ are the polarization and inversion decay
constants, respectively.

In the following, we work in the classical approximation, whereby the noise
forces are neglected and the operators are treated as c-numbers (the prior
notation will be retained). This approximation fails near the lasing
threshold, where the average intensity is smaller than its quantum
fluctuations. The classical version of Eqs.~\eq{langlamb}-\eq{langs} becomes
\aln{
  &\frac d {dt} |a \rangle = - i \hat \Omega |a \rangle + \sum_{p'} | g_{p'}
  \rangle\, \sigma_{p'}, 
  \label{langsemlamb}\\
  &\dot \sigma_p = -\l( i \nu + \gamma_\perp \r) \sigma_p + 2\, \langle g_p | a
  \rangle\, s_p, 
  \label{langsemsigm}\\
  &\dot s_p = \gamma_\parallel \l(S - s_p \r) - \l( \langle a | g_p \rangle\,
  \sigma_p + \text{c.c.} \r).
  \label{langsems}
}
Here, for the compactness of notation, we introduced the \emph{classical}
vectors $|a \rangle = (\ldots, a_\lambda, a_{\lambda + 1}, \ldots)^T$ and $|
g_p \rangle\ = (\ldots, g_{\lambda p}, g_{\lambda + 1, p}, \ldots)^T$ and the
matrix
\eqn{
  \bigl(\hat \Omega \bigr)_{\lambda \lambda'} =   \omega_\lambda
  \delta_{\lambda \lambda'} - i \gamma_{\lambda \lambda'}.
  \label{Omega}
}
Since $\hat \Omega$ is not Hermitian, it has different left and right
eigenbases $\langle L_k |$ and $| R_k \rangle$, respectively. Thus, its
spectral decomposition is of the form
\eqn{
  \hat \Omega = \sum_k |R_k \rangle \l(\Omega_k - i \kappa_k \r) \langle L_k |,
  \label{Omegad}
}
where the eigenvectors are normalized in such a way that $\langle L_k | R_{k'}
\rangle = \delta_{k k'}$ and $\langle R_k| R_k \rangle = 1$, but, in general,
$\langle R_k| R_{k'} \rangle \ne \delta_{k k'}$. $\Omega_k - i \kappa_k$
($\kappa_k > 0$) are the complex eigenfrequencies of the passive \emph{open}
resonator. We expand $|a \rangle = \sum_k \alpha_k |R_k \rangle$, where the
amplitudes
\eqn{
  \alpha_k = \langle L_k | a \rangle,
}
satisfy the equations  
\aln{
  &\dot \alpha_k = - \l(i \Omega_k + \kappa_k \r) \alpha_k  + \sum_{p'} \langle
  L_k | g_{p'} \rangle\, \sigma_{p'}, 
  \label{langsemalphk}\\
  &\dot \sigma_p = -\l( i \nu + \gamma_\perp \r) \sigma_p + 2 \sum_{k'} \langle
  g_p |  R_{k'} \rangle\, \alpha_{k'} s_p, 
  \label{langsemsigmk}\\
  &\dot s_p = \gamma_\parallel \l(S - s_p \r) - \sum_{k'} \l( \langle R_{k'} |
  g_p \rangle\, \alpha_{k'}^* \sigma_p + \text{c.c.} \r)
  \label{langsemsk}
}
following from Eqs.~\eq{langsemlamb}-\eq{langsems}. It is worth emphasizing
again that the open-resonator modes~$k$ are coupled through the interaction
with the atoms, while the closed-resonator modes~$\lambda$, in addition,
interact via the reservoir.

\subsection{Number of lasing modes}

We solve Eqs.~\eq{langsemalphk}-\eq{langsemsk} by treating the interaction with
the atoms perturbatively. Namely, it will be assumed that the sustained field
oscillations are proportional to the unperturbed eigenvectors~$|R_k \rangle$.
However, the oscillation frequencies~$\wbar \Omega_k \approx \Omega_k$ are to
be determined selfconsistently. It was argued in Ref.~\cite{fu91}, that a
multimode solution is possible~if 
\eqn{
  \kappa_k, \gamma_\parallel \ll \Delta \wbar \Omega, \gamma_\perp,  
}
where $\Delta \wbar \Omega$ is a typical spacing between the lasing-mode
frequencies. This condition ensures that the population inversion~$s_p$ is
approximately constant in time~\footnote{The same is tacitly assumed in
Ref.~\cite{hack05}}. An expression for~$s_p$ can be derived as follows. First,
one represents the polarization $\sigma_p = \sum_k \sigma_{pk}$ as a sum of
single-frequency components $\sigma_{pk} \propto \exp\l( -i \wbar \Omega_k t
\r)$. Using Eq.~\eq{langsemsigmk} with $s_p = \text{const}$, 
\eqn{
  \sigma_{pk} = \frac {2 s_p \langle g_p |  R_k \rangle} {-i \l(\wbar \Omega_k
  - \nu \r) + \gamma_\perp}\, \alpha_k
  \label{sigmapk}
}
is expressed in terms of~$\alpha_k$, which oscillates with the same frequency.
Finally, these $\sigma_{pk}$ are substituted to Eq.~\eq{langsemsk} yielding
\aln{
  &s_p \approx \frac S {Y_p}, 
  \label{sp} \\
  &Y_p \equiv 1 + \frac 4 {\gamma_\perp \gamma_\parallel} \sum_k |
  \langle g_p | R_k \rangle|^2 \mc L_k\, |\alpha_k|^2,
  \label{Yp} \\
  &\mc L_k \equiv \l[1 + \l(\wbar \Omega_k - \nu \r)^2 / \gamma_\perp^2
  \r]^{-1},
  \label{Lor}
}
where the oscillating products $\alpha_k^* \alpha_{k'}^{}$, $k \ne k'$, were
averaged out, in line with the constant-$s_p$ approximation. Taking into
account Eqs.~\eq{sigmapk} and~\eq{sp} and keeping only the terms oscillating
with frequency~$\wbar \Omega_k$ in Eq.~\eq{langsemalphk}, we arrive at an
equation for~$\alpha_k$, 
\aln{
  &\dot \alpha_k \approx - \l( i \Omega_k + \kappa_k - B_k \r) \alpha_k, 
  \label{alphk}\\
  &B_k \equiv \frac {2S} {-i \l(\wbar \Omega_k - \nu \r) + \gamma_\perp}\,
  \langle L_k | \sum_p \frac {| g_p \rangle \langle g_p |} {Y_p} | R_k \rangle.
  \label{Bk}
} 
Equations~\eq{alphk} for $N$ lasing modes are equivalent to a system of $2N$
real equations~\cite{hack05},
\aln{
  &\wbar \Omega_k = \Omega_k - \text{Im}\, B_k, 
  \label{freq}\\
  &\text{Re}\, B_k = \kappa_k,
  \label{inteq}
}
from which the frequencies~$\wbar \Omega_k$ and the intensities $I_k =
|\alpha_k|^2$ of the lasing modes can be determined. Equations~\eq{freq}
and~\eq{inteq} are valid only for such~$k$ that $I_k > 0$. 

Further progress can be made if $Y_p^{-1}$ in Eq.~\eq{Bk} is expanded up to the
linear terms in~$I_k$. This procedure presumes that the laser is operating not
far from the threshold. When the atoms are distributed uniformly over the
resonator and their density is sufficiently large, the $p$~sum becomes a volume
integral. Then the summations entering Eq.~\eq{Bk} are computed as follows:
\aln{
  &\sum_p | g_p \rangle \langle g_p | \to \mc N g^2 \int d^3 r\,  | \phi^*
  (\mbf r) \rangle \langle \phi^* (\mbf r) | = \mc N g^2, 
  \label{2prod} \\
  &\sum_p \langle L_k | g_p \rangle \langle g_p | R_k \rangle | \langle g_p |
  R_{k'} \rangle|^2 \notag \\
  &\to \mc N g^4 V \int d^3 r\, \langle L_k | \phi^* (\mbf r) \rangle \langle
  \phi^* (\mbf r) | R_k \rangle | \langle \phi^* (\mbf r) | R_{k'} \rangle|^2
  \notag \\
  &\approx \mc N g^4 (1 + 2 \delta_{kk'}).    
  \label{4prod}
}
where $|\phi (\mbf r) \rangle = \bigl(\ldots, \phi_\lambda (\mbf r),
\phi_{\lambda + 1} (\mbf r), \ldots \bigr)^T$, 
\eqn{
  g = \sqrt{\frac {2 \pi \nu} {\hbar V}}\, |\mbf d_{21}|,
}
and $V$ is the resonator volume. Above we assumed that all the modes~$\lambda$
are polarized along~$\mbf d_{21}$ and $\omega_\lambda \approx \nu$. The
approximate equality in Eq.~\eq{4prod} results from treating the wavefunctions
$\langle L_k | \phi^* (\mbf r) \rangle$ and $\langle \phi^* (\mbf r) | R_k
\rangle$ in a wave-chaotic resonator as Gaussian random variables, restricted
only by normalization. Performing random-matrix simulations, this property was
shown to hold in the relevant range of the mode widths~\cite{hack05}. Using
Eqs.~\eq{2prod} and~\eq{4prod}, we arrive~at
\aln{
  B_k = &\frac {2S \mc N g^2} {\gamma_\perp} \mc L_k \l(1 + i \frac {\wbar
  \Omega_k - \nu} {\gamma_\perp} \r) \notag \\ 
  &\times \l[ 1 - \frac {4 g^2} {\gamma_\perp \gamma_\parallel} \sum_{k'}
  I_{k'} \mc L_{k'} (1 + 2 \delta_{kk'}) \r] + O(I^2).
  \label{Bk_appr}
}
The linear gain~$G_0 \mc L_k$, where 
\eqn{
  G_0 \equiv \frac {2 \mc N g^2} {\gamma_\perp}\, S,
}
is obtained from $\text{Re}\, B_k$ by setting~$I_{k'} = 0$. 

To determine the number of lasing modes, we approximate $\wbar \Omega_k$ with
$\Omega_k$, substitute $B_k$~\eq{Bk_appr} in Eq.~\eq{inteq}, divide it by~$\mc
L_k$, sum over~$k$, and find $\sum_{k'} I_{k'} \mc L_{k'}$. Then $\sum_{k'}
I_{k'} \mc L_{k'}$ can be used in Eq.~\eq{inteq} to express~$I_k$, which is
required to be positive for all lasing modes. This condition
yields~\cite{misi98}
\eqn{
  \frac {\kappa_N} {\mc L_N} + \frac 1 2 \l(N \frac {\kappa_N} {\mc L_N} -
  \sum_{k=1}^N \frac {\kappa_k} {\mc L_k} \r) <  G_0,
  \label{N}
}
where the modes are ordered in such a way that $\kappa_1/ \mc L_1 \leq
\kappa_2/ \mc L_2 \leq \ldots$. The largest $N$ satisfying this inequality is
the number of lasing modes. On the other hand, the largest~$N_0$ such that
\eqn{
  \frac {\kappa_{N_0}} {\mc L_{N_0}} < G_0
  \label{N0}
}
holds, is the number of modes that would lase in the absence of mode
competition. These are the modes for which the linear gain exceeds the losses.
Clearly, ${N_0 \geq N}$.

\section{Average number of lasing modes and its fluctuations}

\subsection{Random-matrix model}

We investigated the mode statistics resulting from Eqs.~\eq{N} and~\eq{N0}. 
Ensembles of open chaotic resonators were modeled by randomly generated 
non-Hermitian matrices~$\hat \Omega \equiv \hat \omega - i \hat
\gamma$~\eq{Omega}. In the basis of the modes~$\lambda$, $\hat \Omega$~has a
diagonal real part. Its imaginary part is of the form~\cite{hack03}
\eqn{
  - \hat \gamma= - \pi \hat W \hat W^\dag,
}
where $\bigl( \hat W \bigr)_{\lambda m}$ ($\lambda = 1, \ldots, L$, $m = 1,
\ldots, M \leq L$) describes interaction of the resonator mode~$\lambda$ with
the $m$th channel of the reservoir. Each channel includes a continuum of
frequencies. However, $\hat W$~is frequency independent in the Markov
approximation. Clearly, the matrix $\hat \gamma$ has at most $M$ nonzero
eigenvalues $\gamma_i$, $i = 1, \ldots, M$ (it is possible to find $N-M$
linearly independent vectors orthogonal to the rows of~$\hat W^\dag$). We will
consider the case of equivalent channels, when all $\gamma_i = \gamma$.

\begin{figure}
  \centering{\includegraphics[width= 0.85\linewidth, angle=0]{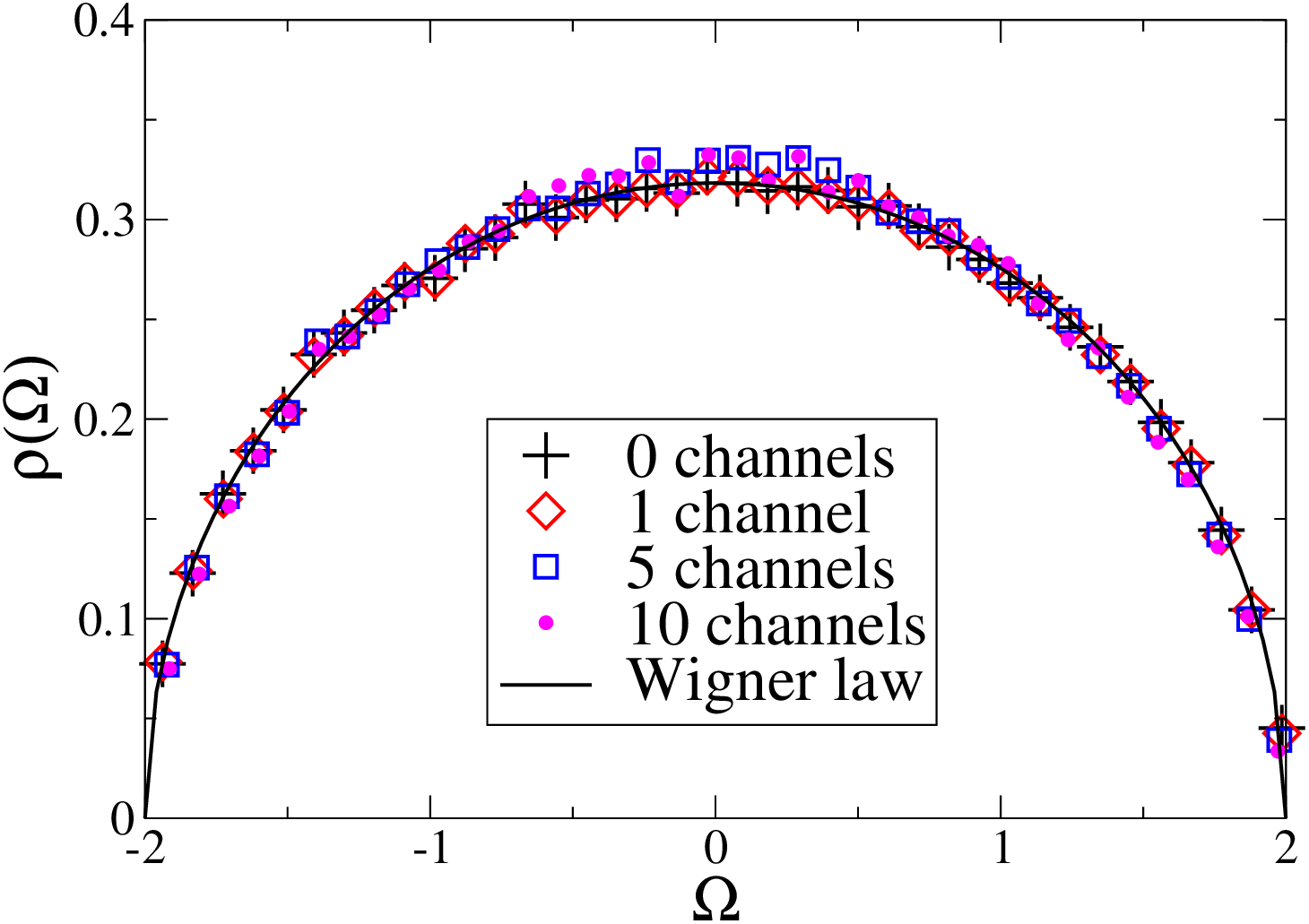}} 
  \caption{(Color online) Distribution $\rho (\Omega)$ of the real parts of
  the eigenvalues of random matrices. Numerical simulations were performed for
  an ensemble of 500 matrices of size $L=200$ coupled to $M = 0$, $1$, $5$,
  and $10$ open channels with the coupling $\gamma = 1$. Analytical
  expression~\eq{wig} is plotted for comparison.
  \vspace*{.5cm}}
  \label{wigner_fig}
\end{figure}

A matrix~$\hat \Omega$ is most easily constructed in the basis where $\hat
\gamma$ is diagonal: we fix~$\hat \gamma$ and choose $\hat \omega$ from a
Gaussian orthogonal ensemble~\cite{somm99}. Without loss of generality, the
diagonal and off-diagonal elements of~$\hat \omega$ are taken from a normal
distribution with zero mean and the variance of $2/L$ and $1/L$, respectively.
In the limit~$L \gg 1$, $\Omega_k$~\eq{Omegad} (the \emph{real} parts of the
eigenvalues of~$\hat \Omega$) are distributed according to the Wigner
semicircle~law
\eqn{
  \rho (\Omega) = \frac 1 \pi \sqrt{1 - \frac {\Omega^2} 4}, \quad -2 \leq
  \Omega \leq 2,
  \label{wig}
}
where $\rho (\Omega)$ is normalized to unity. Numerical simulations
(Fig.~\ref{wigner_fig}) show a reasonable agreement with this equation. The
strength of coupling to continuum is characterized by a parameter $2 \pi \rho
(\Omega) / \l(\gamma + \gamma^{-1} \r)$~\cite{somm99}. Thus, it is sufficient
to consider $\gamma \in [0, 1]$, whereby $\gamma = 0$ ($\gamma = 1$)
corresponds to the vanishing (strongest) coupling. Importantly, even within one
matrix~$\hat \Omega$, the effective coupling depends on the spectral region
according to~$\rho (\Omega)$.

\subsection{Results and discussion}

In the following figures, we present results of numerical simulations for the
average number of lasing modes~$\langle N \rangle$ and its standard
deviation~$\sigma_N$. The respective quantities in the absence of the mode
competition, $\langle N_0 \rangle$ and $\sigma_{N_0}$, were calculated as
well. The averages were performed over ensembles of random matrices. As was
explained earlier, the effective coupling to continuum depends on~$\rho
(\Omega)$. Therefore, for each matrix, of all eigenvalues, only $L_0 \approx
0.36 L$ eigenvalues closest to the top of the Wigner semicircle were used in
Eqs.~\eq{N} and~\eq{N0}. Within this spectral region, $\rho (\Omega)$~varies by
about~4\%. In order to reduce the number of parameters, we assumed that
$\gamma_\perp$ is sufficiently large, and set $\mc L_k = 1$. The pumping was
measured in units of its threshold value~$S_{\text{thr}}$, which was determined
numerically from the threshold condition $\langle N \rangle = 1$. An estimate
yields $S / S_{\text{thr}} \sim G_0 / \kappa_0$, where $\kappa_0 \sim \gamma M
/ L$ is a typical loss. 

\begin{figure}[tb]
  \centering{\includegraphics[width= 0.85\linewidth, angle=0]
  {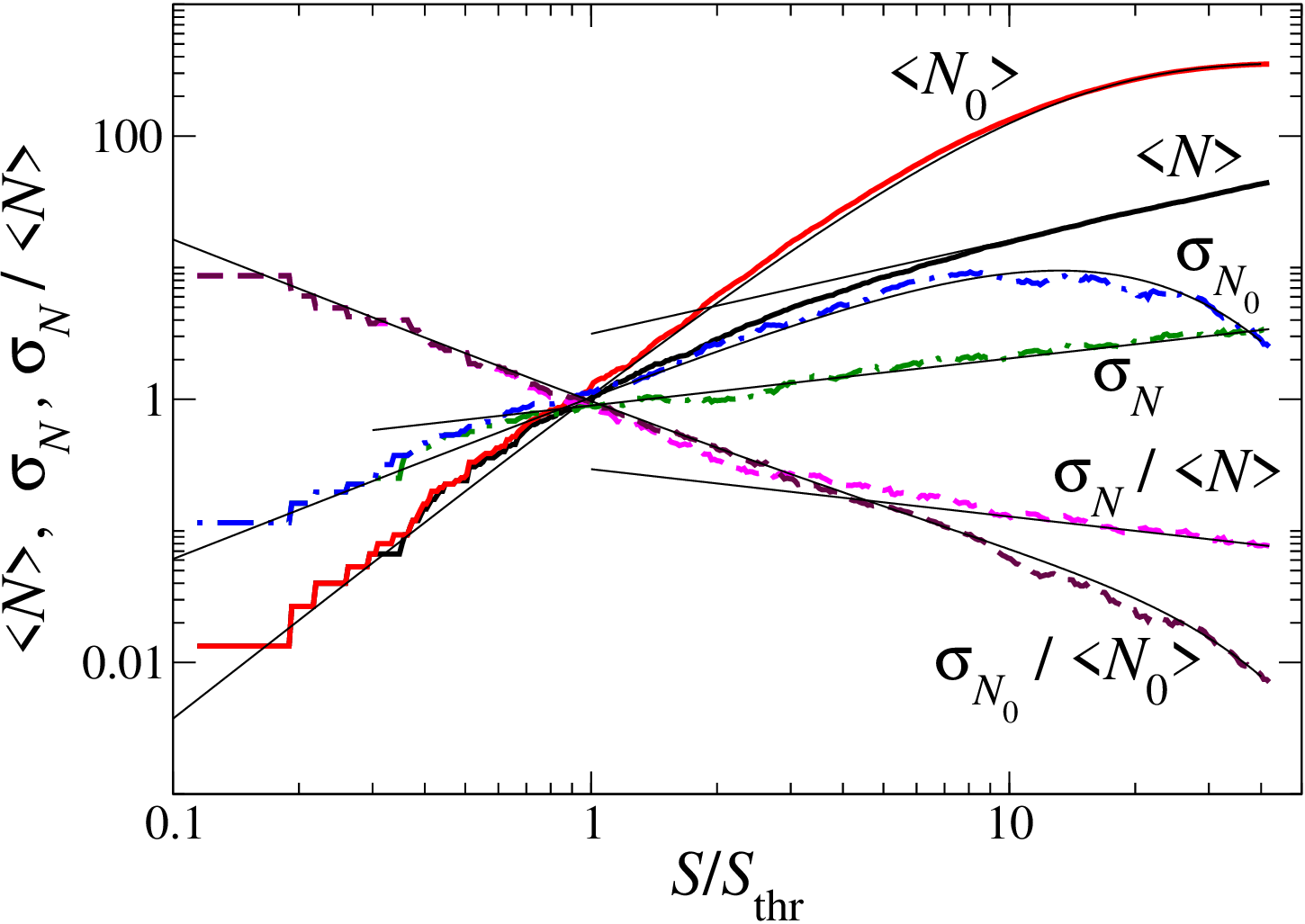}} 
  \caption{(Color online) Average number of lasing modes~$\langle N \rangle$,
  its standard deviation~$\sigma_N$, and relative standard deviation~$\sigma_N
  / \langle N \rangle$, as a function of pumping. The same in the absence of
  mode competition ($\langle N_0 \rangle$, $\sigma_{N_0}$, $\sigma_{N_0} /
  \langle N_0 \rangle$). The thin solid lines display analytical results of
  Eqs.~\eq{N0an} and~\eq{N0sd} and power-law asymptotics. Numerical
  simulations were performed for an ensemble of 75 random matrices of size $L
  = 1000$ coupled to $M = 5$ open channels with the coupling $\gamma = 0.1$.
  $L_0 = 360$ eigenvalues per matrix were taken into account. The analytical
  functions were computed with the parameters $\wbar \kappa = 5.0 \times
  10^{-4}$ and $G_0\l|_{S=S_{\text{thr}}} \r. = 3.3 \times 10^{-5}$, which
  were extracted from the ensemble.
  \vspace*{.5cm}}
  \label{modstat_5_p1_fig}
\end{figure}

First, we discuss a weakly open resonator, $\kappa_0 \ll \Delta \Omega \approx
\pi / L$, where $\Delta \Omega$ is the mean nearest-neighbor spacing
between~$\Omega_k$. A theory for the averages $\langle N \rangle$ and $\langle
N_0 \rangle$ in this regime was proposed in Ref.~\cite{misi98}. Central to the
argument is the analytical expression for the distribution
of~$\kappa_k$~\eq{Omegad},
\eqn{
  P(y) = \frac {(M/2)^{M/2}} {\Gamma (M/2)}\, y^{\frac M 2 - 1} \exp \l( -
  \frac M 2\, y \r), \quad y \equiv \kappa / \wbar \kappa,
  \label{Py}
}
where $\wbar \kappa$ is the average of~$\kappa$ and $\Gamma (z)$ is the gamma
function~\cite{abra72}. $P(y)$~is a $\chi^2$~distribution with $M$ degrees of
freedom and the average $\wbar y = 1$. The average number of modes in the
absence of competition can be estimated from
\eqn{
  \langle N_0 \rangle = L_0 \int_0^{G_0/\wbar \kappa} dy\, P(y) = L_0\, \frac
  {\gamma \l( \frac M 2, \frac M 2 \frac {G_0} {\wbar \kappa} \r)} {\Gamma
  (M/2)},
  \label{N0an}
}
where $\gamma (z, x) = \int_0^x dt\, t^{z-1} e^{-t}$ is the incomplete gamma
function~\cite{abra72}. The numerical results in Fig.~\ref{modstat_5_p1_fig}
show a good agreement with this prediction. In the weak-pump regime $G_0/\wbar
\kappa \ll 1$, there is a power law $\langle N_0 \rangle \propto
S^{M/2}$~\cite{misi98}. With the increased pumping, the saturation $\langle N_0
\rangle \to L_0$ sets~in. Clearly, this form of saturation is an artifact of
our model. Normally, the number of potential lasing modes would be limited by
the Lorentzians~$\mc L_k$. Nevertheless, for large~$L_0$, the present model is
appropriate below the saturation. 

In order to determine~$\sigma_{N_0}$, it is the simplest to assume that $N_0 =
1, \ldots, L_0$ is distributed according to a binomial distribution $P_p (N_0 |
L_0)$, where $p = \langle N_0 \rangle / L_0$ is the probability for a given
mode to lase. The standard deviation for this distribution is known to~be
\eqn{
  \sigma_{N_0} = \sqrt{ \langle N_0 \rangle \left(1 - \frac {\langle N_0
  \rangle} {L_0} \r)},
  \label{N0sd}
}
where $\langle N_0 \rangle$ can be substituted from Eq.~\eq{N0an}. A comparison
with the numerical curve in  Fig.~\ref{modstat_5_p1_fig} supports our
assumption. When $\langle N_0 \rangle \ll L_0$, we find a power law
$\sigma_{N_0} \propto \sqrt{\langle N_0 \rangle} \propto S^{M/4}$. 

In the presence of mode competition, it is interesting to look at the case when
$\langle N \rangle$ is far below the saturation, but, still, sufficiently
large. The former condition yields $\langle N \rangle \propto \kappa^{M/2}$,
while the latter ensures that the terms of order $N \kappa_N \approx \langle N
\rangle \kappa$ dominate the left-hand side of Eq.~\eq{N}. Combination of the
two estimates gives $\langle N \rangle \propto S^{\frac M
{M+2}}$~\cite{misi98}. We checked numerically (via~$\sigma_N$) that the
$N$~distribution is not binomial. Nevertheless, the power law $\sigma_N \propto
\sqrt{\langle N\rangle}\propto S^{\frac M {2(M+2)}}$ remains valid
(Fig.~\ref{modstat_5_p1_fig}). 

\begin{figure}[tb]
  \centering{\includegraphics[width= 0.85\linewidth, angle=0]
  {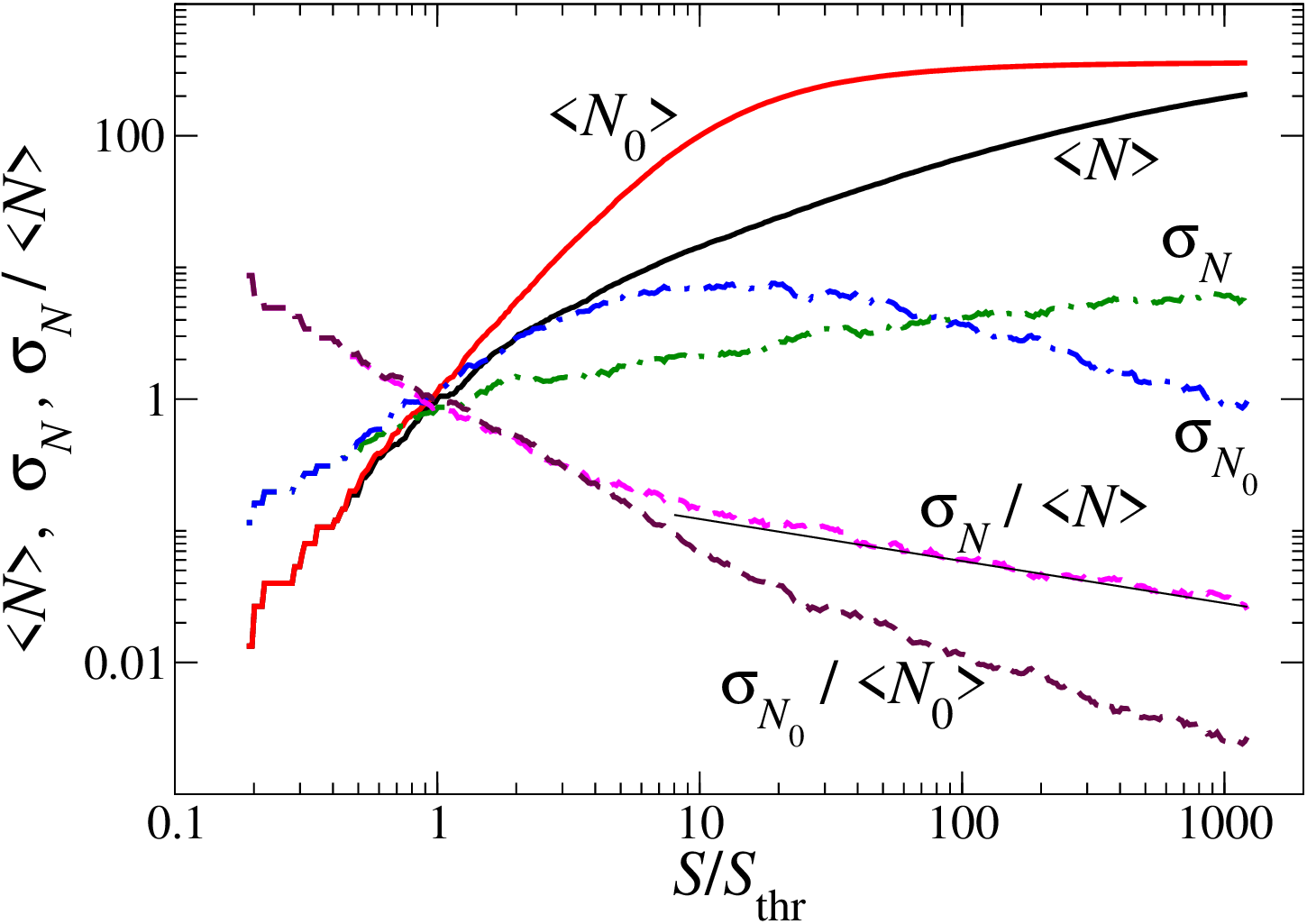}} 
  \caption{(Color online) Same as in Fig.~\ref{modstat_5_p1_fig}, but with the
  coupling $\gamma = 1.0$. The thin solid line displays a power law with the
  exponent~$-0.32$. The ensemble is characterized by $\wbar \kappa = 7.4
  \times 10^{-3}$ and $G_0\l|_{S=S_{\text{thr}}} \r. = 1.1 \times 10^{-4}$.
  \vspace*{.5cm}}
  \label{modstat_5_1p_fig}
\end{figure}

Next, we consider the problem of a strong coupling to the bath ($\kappa_0 \sim
\Delta \Omega$) modeled here by a random-matrix ensemble with $\gamma = 1$. The
distribution $P (y)$ in this case is no longer given by Eq.~\eq{Py}, but rather
has a power-law tail $P (y) \propto y^{-2}$, $y \gg 1$~\cite{somm99}; its full
analytical expression is very complicated. Numerical results for the average
number of lasing modes and its fluctuations were obtained for $M = 1$, $3$,
$5$, $7$, and~$10$. The data for $M=5$ are presented in
Fig.~\ref{modstat_5_1p_fig}. The $S$~dependencies of $\langle N \rangle$ and
$\langle N_0 \rangle$ here are similar to those found in Ref.~\cite{hack05}.
However, except for $M=1$, we can confirm for these quantities neither a
power-law behavior, in general, nor the powers~$1/3$, respectively,~$1/2$, in
particular. 

\begin{figure}[tb]
  \centering{\includegraphics[width= 0.85\linewidth, angle=0]{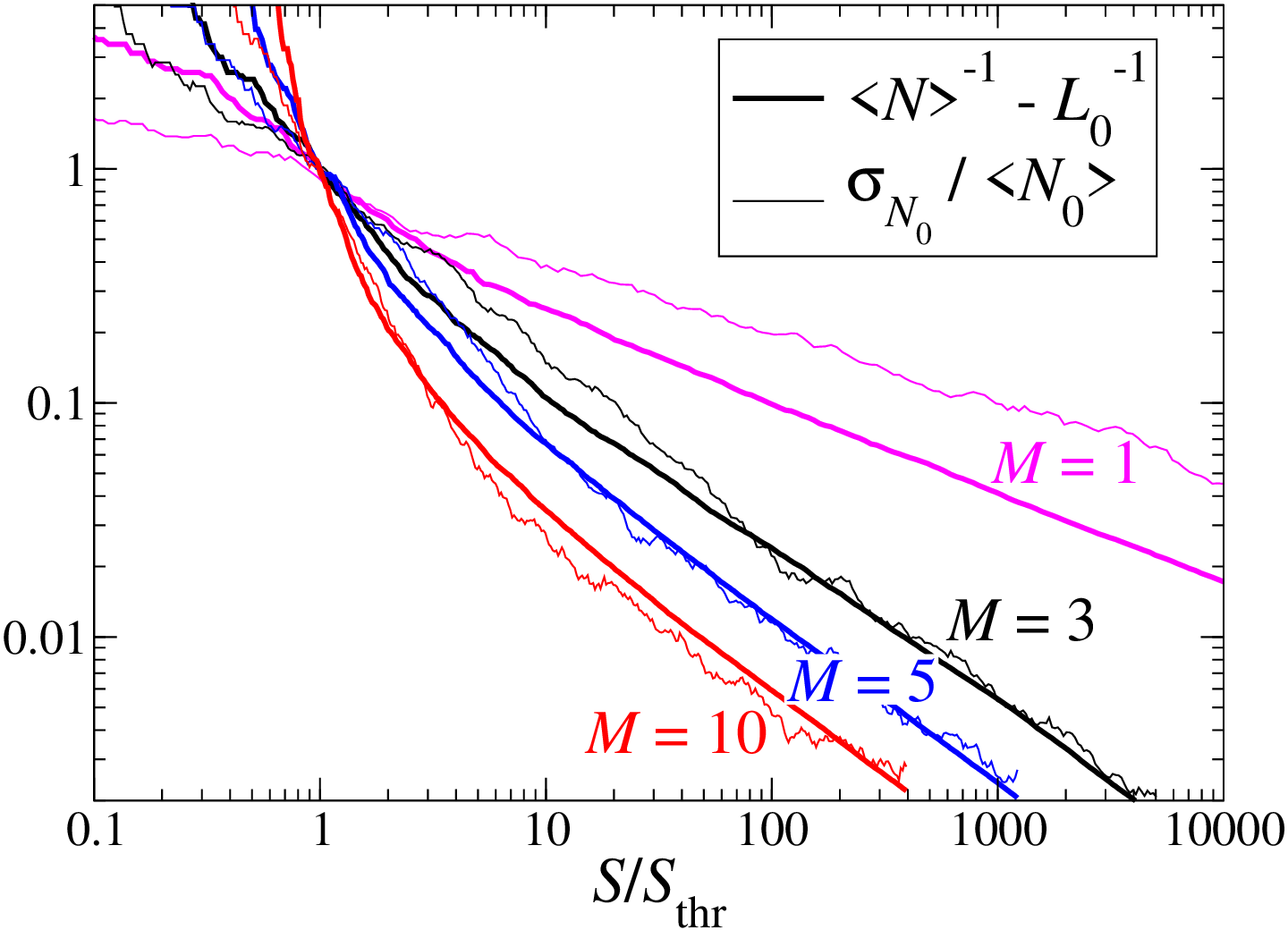}} 
  \caption{(Color online) Comparison of $\langle N \rangle^{-1} -L_0^{-1}$
  (thick lines) and $\sigma_{N_0} / \langle N_0 \rangle$ (thin lines) plotted
  as functions of pumping. The ensembles of random matrices as in
  Fig.~\ref{modstat_5_p1_fig}, but with the coupling $\gamma = 1.0$ and with
  the number of open channels~$M$ labeled in the figure.
  \vspace*{.5cm}}
  \label{relation_fig}
\end{figure}

An analysis of fluctuations shows that the distributions of $N$ and~$N_0$ are
non-binomial. While the standard deviation $\sigma_N$ saturates at large~$S$,
the relative deviation $\sigma_N/ \langle N \rangle$ exhibits a power law with
the exponent $0.2 \div 0.4$, depending on~$M$. Examining the numerical data, we
discovered an interesting relation between the relative fluctuations of~$N_0$
and the average of~$N$:
\eqn{
  \frac {\sigma_{N_0}} {\langle N_0 \rangle} \approx \frac 1 {\langle N
  \rangle} - \frac 1 {L_0}, \quad \frac S {S_{\text{thr}}} \gg 1.
  \label{rel}
}
As demonstrated in Fig.~\ref{relation_fig}, this property is well satisfied for
$M \ne 1$. Unfortunately, an explanation of this result is still lacking. 

\section{Conclusions}

By modeling ensembles of open chaotic resonators with ensembles of random
matrices, we studied average number of lasing modes and its fluctuations. To
highlight the effect of mode competition and allow for a better comparison
with earlier work, the linear approximation to the lasing equations was
considered as well. 

In the case of a weakly open resonator, the average number of modes is
proportional to a power of pump strength (within a certain pumping range).
This result agrees with the analytical prediction in Ref.~\cite{misi98}. The
standard deviation changes as a square root of the average. In the absence of
mode competition, the number of modes follows a binomial distribution, if the
total number of eigenstates available for lasing is finite. The distribution
becomes non-binomial in the presence of mode competition.

For a resonator strongly coupled to the environment, we could not establish any
power-law dependence of the average mode number on pumping. (The case of one
open channel makes an exception.) This evidence stays in contradiction to the
conclusions of Ref.~\cite{hack05}. On the other hand, we find a power-law
behavior for the relative fluctuations. A curious relation~\eq{rel} between the
average with and the relative fluctuation without the mode competition requires
further investigation. 

As a possible extension of this work, it would be interesting to relax some of
the assumptions made. For example, one can consider an effect of the Lorentzian
line shape (here approximated as rectangular). A more challenging task is to
avoid the near-threshold expansion of the intensity-dependent denominators in
Eq.~\eq{Bk}.

\vspace{.5cm}
\begin{acknowledgments}

The author is grateful to Fritz Haake for introducing him to the research area
of random lasers and continuing support of the project. Hui~Cao, Gregor
Hackenbroich, Dmitry Savin, and Hans-J\"urgen Sommers are acknowledged for
helpful discussions. This work was financially supported by the Deutsche
Forschungsgemeinschaft via the SFB ``Transregio~12.''

\end{acknowledgments}

\bibliography{modvar}

\begin{thebibliography}{20}
\expandafter\ifx\csname natexlab\endcsname\relax\def\natexlab#1{#1}\fi
\expandafter\ifx\csname bibnamefont\endcsname\relax
  \def\bibnamefont#1{#1}\fi
\expandafter\ifx\csname bibfnamefont\endcsname\relax
  \def\bibfnamefont#1{#1}\fi
\expandafter\ifx\csname citenamefont\endcsname\relax
  \def\citenamefont#1{#1}\fi
\expandafter\ifx\csname url\endcsname\relax
  \def\url#1{\texttt{#1}}\fi
\expandafter\ifx\csname urlprefix\endcsname\relax\def\urlprefix{URL }\fi
\providecommand{\bibinfo}[2]{#2}
\providecommand{\eprint}[2][]{\url{#2}}

\bibitem[{\citenamefont{Cao}(2003)}]{cao03}
\bibinfo{author}{\bibfnamefont{H.}~\bibnamefont{Cao}}, \bibinfo{journal}{Waves
  Random Media} \textbf{\bibinfo{volume}{13}}, \bibinfo{pages}{R1}
  (\bibinfo{year}{2003}).

\bibitem[{\citenamefont{Sargent~III et~al.}(1974)\citenamefont{Sargent~III,
  Scully, and Lamb}}]{sarg74}
\bibinfo{author}{\bibfnamefont{M.}~\bibnamefont{Sargent~III}},
  \bibinfo{author}{\bibfnamefont{M.~O.} \bibnamefont{Scully}},
  \bibnamefont{and} \bibinfo{author}{\bibfnamefont{W.~E.} \bibnamefont{Lamb},
  \bibfnamefont{Jr.}}, \emph{\bibinfo{title}{Laser Physics}}
  (\bibinfo{publisher}{Addison-Wesley Publ. Co.}, \bibinfo{address}{Reading},
  \bibinfo{year}{1974}).

\bibitem[{\citenamefont{Haken}(1985)}]{hake85}
\bibinfo{author}{\bibfnamefont{H.}~\bibnamefont{Haken}},
  \emph{\bibinfo{title}{Light}}, vol.~\bibinfo{volume}{2}
  (\bibinfo{publisher}{North-Holland Publ. Co.}, \bibinfo{address}{Amsterdam},
  \bibinfo{year}{1985}).

\bibitem[{\citenamefont{Hackenbroich et~al.}(2002)\citenamefont{Hackenbroich,
  Viviescas, and Haake}}]{hack02}
\bibinfo{author}{\bibfnamefont{G.}~\bibnamefont{Hackenbroich}},
  \bibinfo{author}{\bibfnamefont{C.}~\bibnamefont{Viviescas}},
  \bibnamefont{and} \bibinfo{author}{\bibfnamefont{F.}~\bibnamefont{Haake}},
  \bibinfo{journal}{Phys. Rev. Lett.} \textbf{\bibinfo{volume}{89}},
  \bibinfo{pages}{083902} (\bibinfo{year}{2002}).

\bibitem[{\citenamefont{Hackenbroich et~al.}(2003)\citenamefont{Hackenbroich,
  Viviescas, and Haake}}]{hack03}
\bibinfo{author}{\bibfnamefont{G.}~\bibnamefont{Hackenbroich}},
  \bibinfo{author}{\bibfnamefont{C.}~\bibnamefont{Viviescas}},
  \bibnamefont{and} \bibinfo{author}{\bibfnamefont{F.}~\bibnamefont{Haake}},
  \bibinfo{journal}{Phys. Rev. A} \textbf{\bibinfo{volume}{68}},
  \bibinfo{pages}{063805} (\bibinfo{year}{2003}).

\bibitem[{\citenamefont{T{\"u}reci et~al.}()\citenamefont{T{\"u}reci, Stone,
  and Collier}}]{ture06}
\bibinfo{author}{\bibfnamefont{H.}~\bibnamefont{T{\"u}reci}},
  \bibinfo{author}{\bibfnamefont{A.~D.} \bibnamefont{Stone}}, \bibnamefont{and}
  \bibinfo{author}{\bibfnamefont{B.}~\bibnamefont{Collier}},
  \bibinfo{note}{e-print \texttt{arXiv:cond-mat/0605673}, 2006}.

\bibitem[{\citenamefont{Haken and Sauermann}(1963)}]{hake63}
\bibinfo{author}{\bibfnamefont{H.}~\bibnamefont{Haken}} \bibnamefont{and}
  \bibinfo{author}{\bibfnamefont{H.}~\bibnamefont{Sauermann}},
  \bibinfo{journal}{Z. Physik} \textbf{\bibinfo{volume}{173}},
  \bibinfo{pages}{261} (\bibinfo{year}{1963}).

\bibitem[{\citenamefont{Misirpashaev and Beenakker}(1998)}]{misi98}
\bibinfo{author}{\bibfnamefont{T.~S.} \bibnamefont{Misirpashaev}}
  \bibnamefont{and} \bibinfo{author}{\bibfnamefont{C.~W.~J.}
  \bibnamefont{Beenakker}}, \bibinfo{journal}{Phys. Rev. A}
  \textbf{\bibinfo{volume}{57}}, \bibinfo{pages}{2041} (\bibinfo{year}{1998}).

\bibitem[{\citenamefont{Hackenbroich}(2005)}]{hack05}
\bibinfo{author}{\bibfnamefont{G.}~\bibnamefont{Hackenbroich}},
  \bibinfo{journal}{J. Phys. A: Math. Gen.} \textbf{\bibinfo{volume}{38}},
  \bibinfo{pages}{10537} (\bibinfo{year}{2005}).

\bibitem[{\citenamefont{Fyodorov and Sommers}(1997)}]{fyod97}
\bibinfo{author}{\bibfnamefont{Y.~V.} \bibnamefont{Fyodorov}} \bibnamefont{and}
  \bibinfo{author}{\bibfnamefont{H.-J.} \bibnamefont{Sommers}},
  \bibinfo{journal}{J. Math. Phys.} \textbf{\bibinfo{volume}{38}},
  \bibinfo{pages}{1918} (\bibinfo{year}{1997}).

\bibitem[{\citenamefont{Fyodorov and Khoruzhenko}(1999)}]{fyod99}
\bibinfo{author}{\bibfnamefont{Y.~V.} \bibnamefont{Fyodorov}} \bibnamefont{and}
  \bibinfo{author}{\bibfnamefont{B.~A.} \bibnamefont{Khoruzhenko}},
  \bibinfo{journal}{Phys. Rev. Lett.} \textbf{\bibinfo{volume}{83}},
  \bibinfo{pages}{65} (\bibinfo{year}{1999}).

\bibitem[{\citenamefont{Sommers et~al.}(1999)\citenamefont{Sommers, Fyodorov,
  and Titov}}]{somm99}
\bibinfo{author}{\bibfnamefont{H.-J.} \bibnamefont{Sommers}},
  \bibinfo{author}{\bibfnamefont{Y.~V.} \bibnamefont{Fyodorov}},
  \bibnamefont{and} \bibinfo{author}{\bibfnamefont{M.}~\bibnamefont{Titov}},
  \bibinfo{journal}{J. Phys. A: Math. Gen.} \textbf{\bibinfo{volume}{32}},
  \bibinfo{pages}{L77} (\bibinfo{year}{1999}).

\bibitem[{\citenamefont{Fyodorov and Sommers}(2003)}]{fyod03}
\bibinfo{author}{\bibfnamefont{Y.~V.} \bibnamefont{Fyodorov}} \bibnamefont{and}
  \bibinfo{author}{\bibfnamefont{H.-J.} \bibnamefont{Sommers}},
  \bibinfo{journal}{J. Phys. A: Math. Gen.} \textbf{\bibinfo{volume}{36}},
  \bibinfo{pages}{3303} (\bibinfo{year}{2003}).

\bibitem[{\citenamefont{Haken}(1984)}]{hake84}
\bibinfo{author}{\bibfnamefont{H.}~\bibnamefont{Haken}},
  \emph{\bibinfo{title}{Laser Theory}} (\bibinfo{publisher}{Springer-Verlag},
  \bibinfo{address}{Berlin}, \bibinfo{year}{1984}).

\bibitem[{\citenamefont{Viviescas and Hackenbroich}(2003)}]{vivi03}
\bibinfo{author}{\bibfnamefont{C.}~\bibnamefont{Viviescas}} \bibnamefont{and}
  \bibinfo{author}{\bibfnamefont{G.}~\bibnamefont{Hackenbroich}},
  \bibinfo{journal}{Phys. Rev. A} \textbf{\bibinfo{volume}{67}},
  \bibinfo{pages}{013805} (\bibinfo{year}{2003}).

\bibitem[{\citenamefont{Viviescas and Hackenbroich}(2004)}]{vivi04}
\bibinfo{author}{\bibfnamefont{C.}~\bibnamefont{Viviescas}} \bibnamefont{and}
  \bibinfo{author}{\bibfnamefont{G.}~\bibnamefont{Hackenbroich}},
  \bibinfo{journal}{J. Opt. B: Quantum Semiclass. Opt.}
  \textbf{\bibinfo{volume}{6}}, \bibinfo{pages}{211} (\bibinfo{year}{2004}).

\bibitem[{\citenamefont{Walls and Milburn}(1995)}]{wall95}
\bibinfo{author}{\bibfnamefont{D.~F.} \bibnamefont{Walls}} \bibnamefont{and}
  \bibinfo{author}{\bibfnamefont{G.~J.} \bibnamefont{Milburn}},
  \emph{\bibinfo{title}{Quantum Optics}} (\bibinfo{publisher}{Springer-Verlag},
  \bibinfo{address}{Berlin}, \bibinfo{year}{1995}).

\bibitem[{\citenamefont{Berry}(1977)}]{berr77}
\bibinfo{author}{\bibfnamefont{M.~V.} \bibnamefont{Berry}},
  \bibinfo{journal}{J. Phys. A: Math. Gen.} \textbf{\bibinfo{volume}{10}},
  \bibinfo{pages}{2083} (\bibinfo{year}{1977}).

\bibitem[{\citenamefont{Fu and Haken}(1991)}]{fu91}
\bibinfo{author}{\bibfnamefont{H.}~\bibnamefont{Fu}} \bibnamefont{and}
  \bibinfo{author}{\bibfnamefont{H.}~\bibnamefont{Haken}},
  \bibinfo{journal}{Phys. Rev. A} \textbf{\bibinfo{volume}{43}},
  \bibinfo{pages}{2446} (\bibinfo{year}{1991}).

\bibitem[{\citenamefont{Abramowitz and Stegun}(1972)}]{abra72}
\bibinfo{editor}{\bibfnamefont{M.}~\bibnamefont{Abramowitz}} \bibnamefont{and}
  \bibinfo{editor}{\bibfnamefont{I.~A.} \bibnamefont{Stegun}}, eds.,
  \emph{\bibinfo{title}{Handbook of Mathematical Functions}}
  (\bibinfo{publisher}{Dover}, \bibinfo{address}{New York},
  \bibinfo{year}{1972}).

\end{thebibliography}

\end{document}